\documentclass[12pt,preprint,usenatbib]{aastex}
\usepackage{graphicx,multirow,footnote,longtable,color}
\DeclareGraphicsExtensions{.pdf,.png,.eps}

\shorttitle{SHUCS II. Star Cluster Population of NGC~2997}
\shortauthors{Ryon et al.}

\begin{document}

\title{The Snapshot Hubble U-Band Cluster Survey (SHUCS) II. Star Cluster Population of NGC~2997}

\author{J. E. Ryon}
\affil{Department of Astronomy, University of Wisconsin-Madison, 475 N. Charter St., Madison, WI, 53706, USA}
\email{ryon@astro.wisc.edu}

\author{A. Adamo}
\affil{Max Planck Institut f\"{u}r Astronomie, K\"{o}nigstuhl 17 D-69117 Heidelberg, Germany}
\affil{The Oskar Klein Centre, Department of Astronomy, AlbaNova, Stockholm University, SE-106 91 Stockholm, Sweden}

\author{N. Bastian}
\affil{Astrophysics Research Institute, Liverpool John Moores University, 146 Brownlow Hill, Liverpool L3 5RF, UK}

\author{L. J. Smith}
\affil{Space Telescope Science Institute and European Space Agency, 3700 San Martin Drive, Baltimore, MD 21218, USA}

\author{J. S. Gallagher III}
\affil{Department of Astronomy, University of Wisconsin-Madison, 475 N. Charter St., Madison, WI, 53706, USA}

\author{I. S. Konstantopoulos}
\affil{Australian Astronomical Observatory, PO Box 915, North Ryde NSW 1670, Australia}

\author{S. Larsen}
\affil{Department of Astrophysics / IMAPP, Radboud University Nijmegen, P.O. Box 9010, 6500 GL Nijmegen, The Netherlands}

\author{E. Silva-Villa}
\affil{(CRAQ) Universit\'{e} Laval, 1045, Avenue de la M\'{e}decine, G1V 0A6 Qu\'{e}bec, Canada}

\author{E. Zackrisson}
\affil{The Oskar Klein Centre, Department of Astronomy, AlbaNova, Stockholm University, SE-106 91 Stockholm, Sweden}

\begin{abstract}
We study the star cluster population of NGC~2997, a giant spiral galaxy located at 9.5~Mpc and targeted by the Snapshot \textit{Hubble} $U$-band Cluster Survey (SHUCS). Combining our $U$-band imaging from SHUCS with archival $BVI$ imaging from $HST$, we select a high confidence sample of clusters in the circumnuclear ring and disk through a combination of automatic detection procedures and visual inspection. The cluster luminosity functions in all four filters can be approximated by power-laws with indices of $-1.7$ to $-2.3$. Some deviations from pure power-law shape are observed, hinting at the presence of a high-mass truncation in the cluster mass function. However, upon inspection of the cluster mass function, we find it is consistent with a pure power-law of index $-2.2\pm0.2$ despite a slight bend at $\sim$$2.5\times10^{4}$~M$_{\odot}$. No statistically significant truncation is observed. From the cluster age distributions, we find a low rate of disruption ($\zeta\sim-0.1$) in both the disk and circumnuclear ring. Finally, we estimate the cluster formation efficiency ($\Gamma$) over the last 100~Myr in each region, finding $7\pm2$\% for the disk, $12\pm4$\% for the circumnuclear ring, and $10\pm3$\% for the entire $UBVI$ footprint. This study highlights the need for wide-field $UBVI$ coverage of galaxies to study cluster populations in detail, though a small sample of clusters can provide significant insight into the characteristics of the population. 
\end{abstract}

\section{Introduction}
\label{intro}

Star formation in galaxies commonly results in the production of star clusters. Star clusters were originally thought to fall into two distinct categories, open and globular, which are separated by vast differences in age and mass. Over the last few decades, studies using observations from the {\it Hubble Space Telescope} ({\it HST}) have discovered large populations of young and intermediate-age clusters spanning several orders of magnitude in mass in nearby galaxies \citep[see reviews of][]{whitmore2003,larsen2006a}. This suggests that in reality, the distribution of star cluster ages and masses is more continuous than bimodal. Whether young massive clusters will survive to become a new generation of globular clusters is still an open question.

Young star clusters are potentially very useful for tracing the stellar populations and star formation properties of their host galaxies. Due to their brightness, young clusters can be detected to much greater distances than individual stars, out to several tens of Mpc \citep[e.g.,][]{adamo2010a, fedotov2011}. A fraction of them may also be long-lived and, like globular clusters, survive for nearly a Hubble time. Because star clusters can be approximated as simple stellar populations (SSPs) \citep[e.g.,][]{bastian2013,cabreraziri2014}, it is more straightforward to determine their properties, such as age, mass, and extinction, than it is for the mixed populations making up the diffuse stellar component of galaxies. 

There are many unanswered questions concerning the properties of young star clusters. One of these is the true form of the cluster mass function (CMF). The CMF is a fundamental property of star cluster populations, and has been observed to take the form of a power-law, $dN/dM \propto M^{\beta}$, with $\beta\approx-2$, over a large range of cluster mass \citep[e.g.,][]{zhang1999,bik2003}. The cluster luminosity function (CLF) also appears to be well-described by a power-law of index $\sim$$-2$, and is more readily-observed than the CMF \citep{larsen2002}. It is difficult to directly relate the CLF to the CMF because the age of a cluster plays an important role in determining its luminosity, but the CLF can reflect the shape of the CMF \citep{gieles2006b}. In fact, bends at the high luminosity end of the CLF led \cite{gieles2006b} to suggest that the CMF may be better described by a Schechter function, $dN/dM \propto M^{\beta}\exp(-M/M_c)$, where $M_c$ is the characteristic `Schechter mass' at which the number of clusters drops off rapidly. This form of the CMF has been confirmed by some observations \citep[e.g.,][]{gieles2006a,bastian2012a}, although some studies have found a pure power-law CMF with index $\sim$$-2$ \citep{chandar2010, whitmore2010}. Other studies have also shown environmental dependence in the form of the CMF \citep[e.g.,][]{konstantopoulos2013} and the value of the Schechter mass (e.g., $M_c\sim2\times10^5$~M$_{\odot}$ for a sample of spirals, \citealt{larsen2009}; $M_c\sim8\times10^5$~M$_{\odot}$ for the Antennae, \citealt{jordan2007}).

Another topic of interest is the rate of cluster disruption. \cite{lada2003} found that less than 4-7\% of embedded clusters end up as bound clusters of age $\sim$100 Myr or greater. Rather than providing evidence of high rates of cluster disruption at young ages, this may simply be due to a small fraction ($<26\%$) of stars being born in dense, bound clusters \citep{bressert2010}. This implies that `infant mortality', which describes the unbinding of a cluster upon expulsion of leftover gas from star formation \citep{lada2003}, may not strongly affect young clusters. On the other hand, \cite{gieles2012} suggested that the fraction of stars formed in bound clusters is actually unknown. They showed that a wide range of initial clustering properties can produce the observed distribution of surface densities of young stellar objects, which is the typical method for studying stellar clustering locally in the Galaxy. At older ages, cluster disruption is dominated by two-body relaxation, stellar evolution, and tidal effects of the host galaxy's potential \citep[e.g.,][]{bastian2008a}. 

The effects of cluster formation and disruption are imprinted in the shape of the cluster age distribution. In most studies, the cluster formation rate is assumed to be roughly constant over time periods of $\sim$1~Gyr, which is reasonable for typical spiral galaxies \citep[e.g.,][]{clarke2006}. In this case, only cluster disruption should affect the shape of the age distribution. This shape is predicted to be a power-law if cluster disruption is independent of cluster mass \citep[e.g.,][]{fall2005}, $dN/dt \propto t_{\mathrm{age}}^{\zeta}$. In contrast, the distribution should have a continously changing power-law index if disruption is dependent on mass \citep[e.g.,][]{lamers2005}.  These predictions have proven difficult to test in extragalactic cluster populations \citep{bastian2012a}. In addition, if cluster disruption depends on the environment of the host galaxy, then the shape of the age distribution will also vary. \cite{bastian2012a} and \cite{silvavilla2014} have observed this effect across the disk of M83. These studies found steeper age distributions, and therefore stronger cluster disruption, at small galactocentric radii. In particular, \cite{silvavilla2014} found $\zeta\sim-0.1$ in two outer fields and $\zeta\sim-0.6$ to $-0.5$ in three inner fields of M83 over an age range of 3 to 300~Myr. The authors suggest these differences are due to stronger tidal fields and a higher surface density of giant molecular clouds causing more cluster disruption, and therefore steeper age distributions, in the inner fields as compared to the outer fields.

Recent studies have shown that the fraction of stars that form in bound clusters, or the cluster formation efficiency, also called $\Gamma$, depends on the star formation rate (SFR) density, $\Sigma_{\mathrm{SFR}}\equiv\mathrm{SFR}/A$, where $A$ is the projected area of the star-forming region. \cite{goddard2010} found a power-law $\Gamma$-$\Sigma_{\mathrm{SFR}}$ relationship wherein systems with higher SFR densities have a higher fraction of stars ending up in bound clusters. Since that work, observations of several nearby galaxies have further confirmed this relation \citep{adamo2011, annibali2011, silvavilla2011, cook2012, silvavilla2013}. This has strong implications for the dependence of star and cluster formation properties on environment, and for determining the star formation histories of galaxies. \cite{kruijssen2012a} derived an analytic theory of $\Gamma$ to describe the connection between local star-forming regions and galaxy-scale environmental properties. His model predicts that locally, short freefall times in the highest-density regions of the interstellar medium (ISM) result in high star formation efficiencies (SFEs) and the formation of dense, bound star clusters. The hierarchical structure of the turbulent ISM naturally allows these high-density regions to form. The density spectrum of the ISM can be predicted using galaxy-scale (or local) observables. Therefore, the expected value for $\Gamma$ can be calculated in a straightforward way. A comparison of observed values of $\Gamma$ and those predicted by this model shows strong agreement.

To further investigate these findings, we designed SHUCS, the Snapshot \textit{Hubble} $U$-band Cluster Survey (GO 12229, PI Linda Smith; see \citealt{konstantopoulos2013}). SHUCS is intended to characterize the star cluster populations of a sample of nearby spiral galaxies. We obtained $U$-band images of these galaxies using the Wide Field Camera 3 (WFC3) on \textit{HST}. The $U$-band data complement existing $BVI$ imaging and help to alleviate the age-extinction degeneracy of optical colors \citep{anders2004}. This allows us to measure accurate ages, masses, and extinctions for large numbers of star clusters in each galaxy to study the form of the age and mass distributions, as well as the fraction of stars forming in bound clusters, and how all of these depend on environment.

In this study, we characterize the cluster population of NGC~2997, a nearby, relatively isolated SAB(rs)c galaxy \citep{devaucouleurs1991} with an inclination of $I\sim45^{\circ}$ \citep{jungwiert1997}. We adopt the distance modulus of \cite{larsen1999a} for NGC~2997, 29.9~mag, to allow for a direct comparison of cluster properties between this study and that work. This distance modulus corresponds to a physical distance of 9.5~Mpc. NGC~2997 is the brightest member of a loose group of galaxies which includes a number of low-mass companions \citep{garcia1993}. Visually, NGC~2997 does not appear disturbed, indicating no obvious recent interactions have taken place. 

This paper is organized as follows. We describe our \textit{HST} observations and data reduction in Section~\ref{obs-data}. In Section~\ref{phot}, we discuss our photometry, cluster selection criteria, and detection limits. In Section~\ref{clusterprop}, we discuss the properties of the cluster population of NGC~2997, including the age and mass distributions and cluster formation efficiency. Finally, we summarize our results in Section~\ref{conclusions}.

\section{Observations \& Data Reduction}
\label{obs-data}

\subsection{\textit{HST} Observations}
\label{hstobs}

The \textit{HST} observations of NGC~2997 are shown in Table~\ref{observations}. All data were retrieved from the Mikulski Archive for Space Telescopes (MAST). We used \textit{Astrodrizzle} to correct for geometric distortion and drizzle the images\footnote{http://www.stsci.edu/hst/HST\_overview/drizzlepac} \citep{fruchter2010, gonzaga2012}.

Three dithered images were obtained from WFC3/UVIS. These suffered from charge transfer efficiency (CTE) degradation. We ran the standalone FORTRAN program provided by the Space Telescope Science Institute\footnote{http://www.stsci.edu/hst/wfc3/documents/newsletters/STAN\_03\_14\_2013\#alpha} to reverse the CTE trail effects in the raw dithered images. The corrected images were drizzled to the native pixel scale of 0.04''.

The ACS/HRC and WFPC2 datasets were undithered, and therefore suffered from residual cosmic rays after drizzling. We used the python version of LaCOSMIC\footnote{http://obswww.unige.ch/$\sim$tewes/cosmics\_dot\_py/} \citep{vandokkum2001} to remove cosmic rays from the undrizzled frames, and then drizzled them. The ACS/HRC data were drizzled to the native pixel scale of $\sim$0.028$\times$0.025''. Each WFPC2 chip was drizzled separately, resulting in native pixel scales of 0.0455" (PC) and 0.0996" (WF2, 3, 4).  We did not correct the ACS/HRC and WFPC2 images for CTE degradation. The ACS/HRC data were taken less than two years after the instrument was installed, when CTE losses were still insignificant \citep{ubeda2012}. The WFPC2 data have no visible CTE trails due to higher background levels than the WFC3 \textit{U}-band image. Our detection limits are quite high in these data (see Section~\ref{detlims}), so any CTE correction would be negligible for bright sources.

\begin{table}
\begin{center}
\caption{NGC~2997 HST Observations \label{observations}}
\begin{tabular}{cccccc}
\tableline
 & & Date & Exp. & Program & Zeropoint\\
Camera & Filter & & Time (s) & ID & (Vega mag)\\
\tableline
WFC3/UVIS & F336W & 2010 Oct 28 & 1800 & 12229 & 23.4836\\
\tableline
\multirow{3}{*}{WFPC2} & F450W & 2001 Aug 02 & 460 & 9042 & 21.987, 22.007, 22.016, 21.984\tablenotemark{a} \\
 & F606W & 1994 June 21 & 160 & 5446 & 22.887, 22.919, 22.896, 22.880\tablenotemark{a} \\
 & F814W & 2001 Aug 02 & 460 & 9042 & 21.639, 21.665, 21.659, 21.641\tablenotemark{a} \\
\tableline
\multirow{4}{*}{ACS/HRC} & F220W & \multirow{4}{*}{2003 Nov 18} & 2000 & 9989 & 21.88 \\
& F330W & & 2000 & 9989 & 22.92\\
& F555W & & 640 & 9989 & 25.27\\
& F814W & & 540 & 9989 & 24.86\\
\tableline
\end{tabular}
\tablenotetext{a}{Each WFPC2 chip has a different zeropoint. The order in which they are listed here is PC, WF2, WF3, and WF4.}
\end{center}
\end{table}

\begin{figure}
\centering
\includegraphics[width=\textwidth]{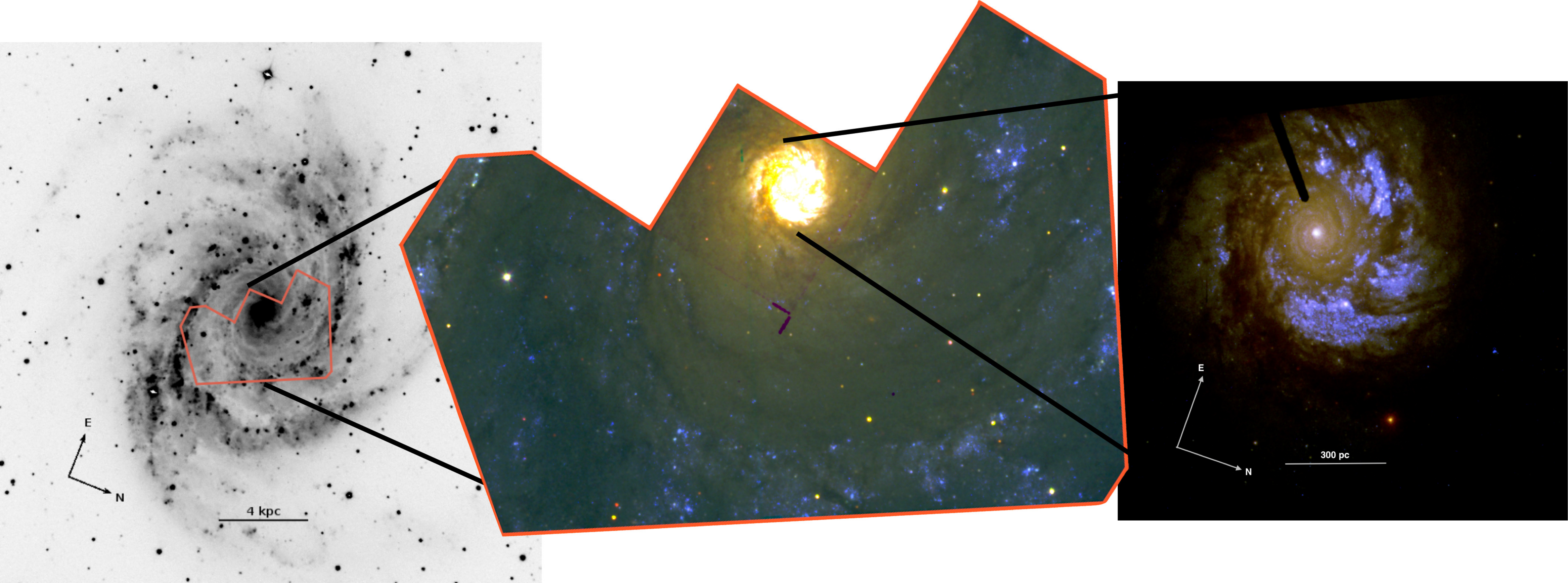}
\caption{Images of NGC~2997. (left) B band image from \cite{larsen1999a} overlaid with an outline showing the $UBVI$ footprint of the \textit{HST} data, (middle) color composite image of the $UBVI$ footprint using the UVIS/F336W, WFPC2/F606W, and WFPC2/F814W images, (right) color composite image of the circumnuclear ring using the HRC/F220W, HRC/F330W, HRC/F555W, and HRC/F814W images. Traces of chip gaps can be seen in the UVIS and WFPC2 composite, as can the occulting finger in the HRC composite. \label{uvi-image}}
\end{figure}

The ACS/HRC data cover the bright circumnuclear ring of NGC~2997. Because the WFPC2 and WFC3/UVIS data were taken at different epochs, they are slightly rotated and offset from one another. Those data cover the circumnuclear ring and a portion of the disk of NGC~2997, and the region over which they overlap contains sections of both spiral arms, as shown in Figure~\ref{uvi-image}. 

In order to take advantage of the high resolution ACS data, we chose to use the UVIS/F336W ($U_{336}$), WFPC2/F450W ($B_{450}$), HRC/F555W ($V_{555}$), and HRC/F814W ($I_{\mathrm{HRC}/814}$) images to complete the $UBVI$ baseline for clusters in the circumnuclear ring. We use the UVIS/F336W ($U_{336}$), WFPC2/F450W ($B_{450}$), WFPC2/F606W ($V_{606}$), and WFPC2/F814W ($I_{\mathrm{WF}/814}$) images to study the clusters in the rest of the disk. We therefore separate our sample into `circumnuclear' and `disk' samples. Circumnuclear sources lie within a 10'' (460~pc) radius from the center of the galaxy, and disk sources lie beyond this radius but within the $UBVI$ footprint.

\subsection{Ancillary Data}
\label{ancillary}

In addition to our \textit{HST} data, we utilized ground-based optical imaging as well as archival infrared (IR) and far-ultraviolet (FUV) imaging from space-based observatories. Optical imaging in $R$ and narrowband H$\alpha$ from the Danish 1.54-m telescope were obtained from the NASA Extragalactic Database (NED). These observations were presented in \cite{larsen1999a}, so we refer the reader to that paper for details on the observations and data reduction. We produced a continuum-subtracted H$\alpha$ image by scaling the $R$ band image using the relative fluxes of isolated stars in both images and then subtracting the scaled $R$ band image from the H$\alpha$ image. We did not flux calibrate the continuum-subtracted H$\alpha$ image because we were not aiming to quantify the strength of H$\alpha$ emission. In Section~\ref{halpha}, we describe the use of these data for constraining the range of models to describe our star clusters.

We obtained post-basic calibration data 24~$\mu$m imaging from MIPS on the \textit{Spitzer} Space Telescope via the NASA/IPAC Infrared Science Archive. We converted this image from surface brightness units (MJy~sr$^{-1}$) to flux units (MJy). We also obtained FUV imaging from the \textit{Galaxy Evolution Explorer} (\textit{GALEX}) via the GalexView tool. We subtracted the sky background map from the FUV image to obtain a sky-subtracted intensity image.  In Section~\ref{cfe}, we describe the photometry performed on these data to calculate the SFR in NGC~2997.

\section{Photometry \& Source Selection}
\label{phot}

\subsection{Photometry Pipeline}
\label{pipeline}

We developed an IRAF pipeline for source selection and photometry, which is described in detail in Section~3 of \cite{konstantopoulos2013}. The pipeline was updated for use with the NGC~2997 dataset, as is described here. We use the $U_{336}$ image as the reference frame for photometry and source selection for two reasons: SHUCS is a $U$-band based survey, and the $U_{336}$ image is the deepest of the data available.

\begin{itemize}

\item[\textbf{1.}] \textbf{Source selection in $U_{336}$.} We used \texttt{daofind} to detect sources on the $U_{336}$ frame with a  detection threshold of 8 times the standard deviation of the background, $\sigma$. We also set the `roundness' and `sharpness' parameters to their full range so as to not bias the selection against elliptical or compact clusters.
\item[\textbf{2.}] \textbf{Coordinate transformation with \texttt{geomap} and \texttt{geoxytran}.}  We use the $U_{336}$ frame as our reference coordinate system. Prior to running the pipeline, we selected 10 to 20 bright reference sources in each image and used \texttt{geomap} to calculate transformation coefficients between $U_{336}$ and the other bands to within rms errors of 0.1 to 0.2 pixels. Within the pipeline, we ran \texttt{geoxytran} on the source catalogue to transform the coordinates of sources detected in step 1. This ensures that we will obtain accurate photometry of identical sources in all bands. 
\item[\textbf{3.}] \textbf{Photometry with \texttt{apphot}.} We use the Vega magnitude system (our photometric zeropoints are listed in Table~\ref{observations}). Photometric apertures of radius 0.16'' were placed at the coordinates on each image defined in step 2. This aperture corresponds to a physical radius of 7.4~pc, which is sufficiently larger than the effective radius ($r_{\mathrm{eff}}$) of a typical star cluster, $\sim$3~pc. Background annuli were placed at a radius of 0.2'' with a width of 1 pixel. We did not allow \texttt{apphot} to recenter the source coordinates.

\end{itemize}

We then calculated aperture corrections from artificial clusters inserted into each image and applied them to our photometry. Empirical stellar PSFs were built from bright, isolated stars in those images where a sufficient number were found. For the remaining images, we used \texttt{Tinytim} to model the PSF \citep{krist2011}. We used the tasks \texttt{MKCMPPSF} and \texttt{MKSYNTH} within \texttt{baolab} \citep{larsen1999b} to create grids of artificial clusters with an \cite{elson1987} light profile and $r_{\mathrm{eff}}=3$~pc. We calculated the aperture corrections by finding the average difference in magnitude between our science photometric aperture of 0.16'', with the sky annulus at 0.2'' of width 1 pixel, and an aperture of 0.8'', with the sky annulus at 0.84'' of width 1 pixel, for each grid of artificial clusters.

Finally, we corrected our photometry for foreground galactic extinction corresponding to $E(B-V)=0.096$ in the direction of NGC~2997, as tabulated by NED. We also added a factor of 0.05~mag in quadrature to our photometric errors to account for zeropoint uncertainties.

\subsection{Concentration Index}
\label{CI}

\begin{figure}
\centering
\includegraphics[scale=0.6]{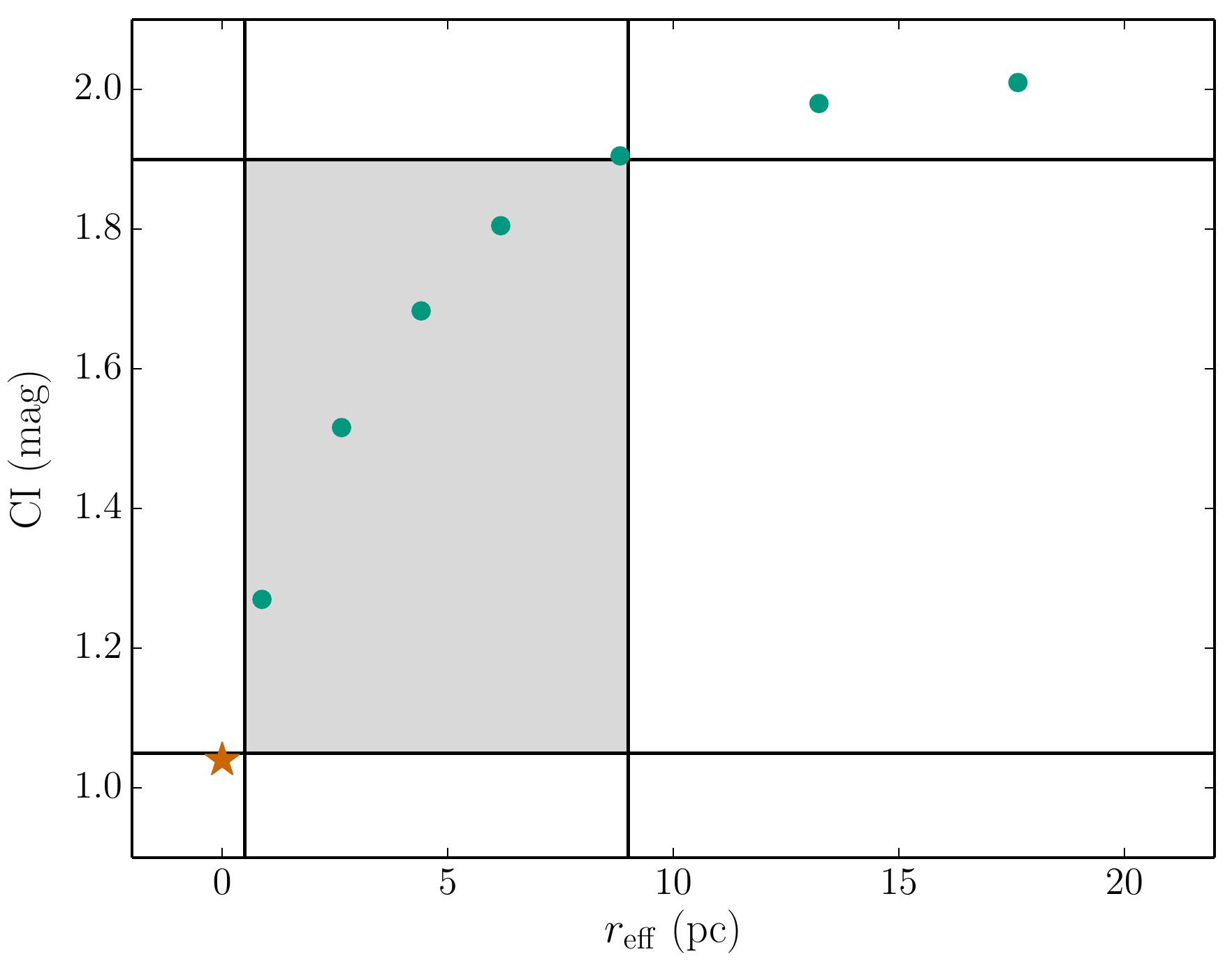}
\caption{Concentration index values over a range of effective radii. The red star shows the average CI for a sample of isolated foreground stars. The blue points represent artificial clusters of various effective radii. The hashed regions denote the range of observed effective radii for star clusters and the corresponding CI values. \label{CI_Reff}}
\end{figure}

The concentration index (CI) is one method for distinguishing between individual stars, star clusters, and stellar associations or galaxies \citep[e.g.,][]{holtzman1996,whitmore2010}. Our concentration index (CI) is the magnitude difference between aperture radii of 1 and 3 pixels in $U_{336}$, and is therefore larger for more extended objects. In order to define the range of acceptable CI values for star clusters at the distance of NGC~2997, we generated grids of artificial clusters for several effective radii using \texttt{baolab} \citep{larsen1999b}. Figure~\ref{CI_Reff} shows the measured CI values for artificial clusters of different effective radii. The average CI value for a sample of isolated foreground stars was 1.04~mag, so we chose a lower limit of 1.05~mag for star clusters. Our upper limit of 1.90~mag corresponds to a physical size of 9~pc, the typical upper effective radius of clusters \citep{portegieszwart2010}.

\subsection{Final Star Cluster Selection Criteria \& Visual Inspection}
\label{selection}

We applied several criteria to our source catalogue to select candidate star clusters. Each source is assigned an $f_{UBVI}$ flag, which defines the criteria it has passed. The possible values of $f_{UBVI}$ are described here:

\begin{itemize}
\item[1:] must be detected and photometered in $UBVI$.
\item[11:] includes above and must have photometric errors $<0.3$~mag.
\item[111:] includes above and must have 1.05~mag $<$ CI $<$ 1.90~mag.
\end{itemize}

In order to build a clean sample of bona-fide clusters, we also performed a visual inspection of sources with $f_{UBVI}=111$ in the $U_{336}$ image. We ran \texttt{ishape} on those sources with $f_{UBVI}=111$, and separated them into three samples based on the $r_{\mathrm{eff}}$ and $S/N$ resulting from the \texttt{ishape} fits. The `good' sample had $0.5$~pc $\leq r_{\mathrm{eff}} \leq 10$~pc and $S/N \geq 30$, while the `doubtful' sample had $r_{\mathrm{eff}}$ outside this range and $S/N \geq 30$. The `uncertain' sample was made up of sources for which the \texttt{ishape} fits resulted in an error, as well as those with $S/N < 30$. Each sample was inspected visually. We assigned an $f_{\mathrm{vis}}$ flag to each source based on the following criteria:

\begin{itemize}
\item[0:] Not extended relative to the PSF, noncircular, or several peaks in intensity due to crowding or confusion.
\item[1:] Clearly extended relative to the PSF, centrally concentrated, and a single peak in intensity. 
\item[2:] Slightly extended relative to the PSF, somewhat circular, and a single peak in intensity.
\end{itemize}

\begin{figure}
\centering
\includegraphics[width=\textwidth]{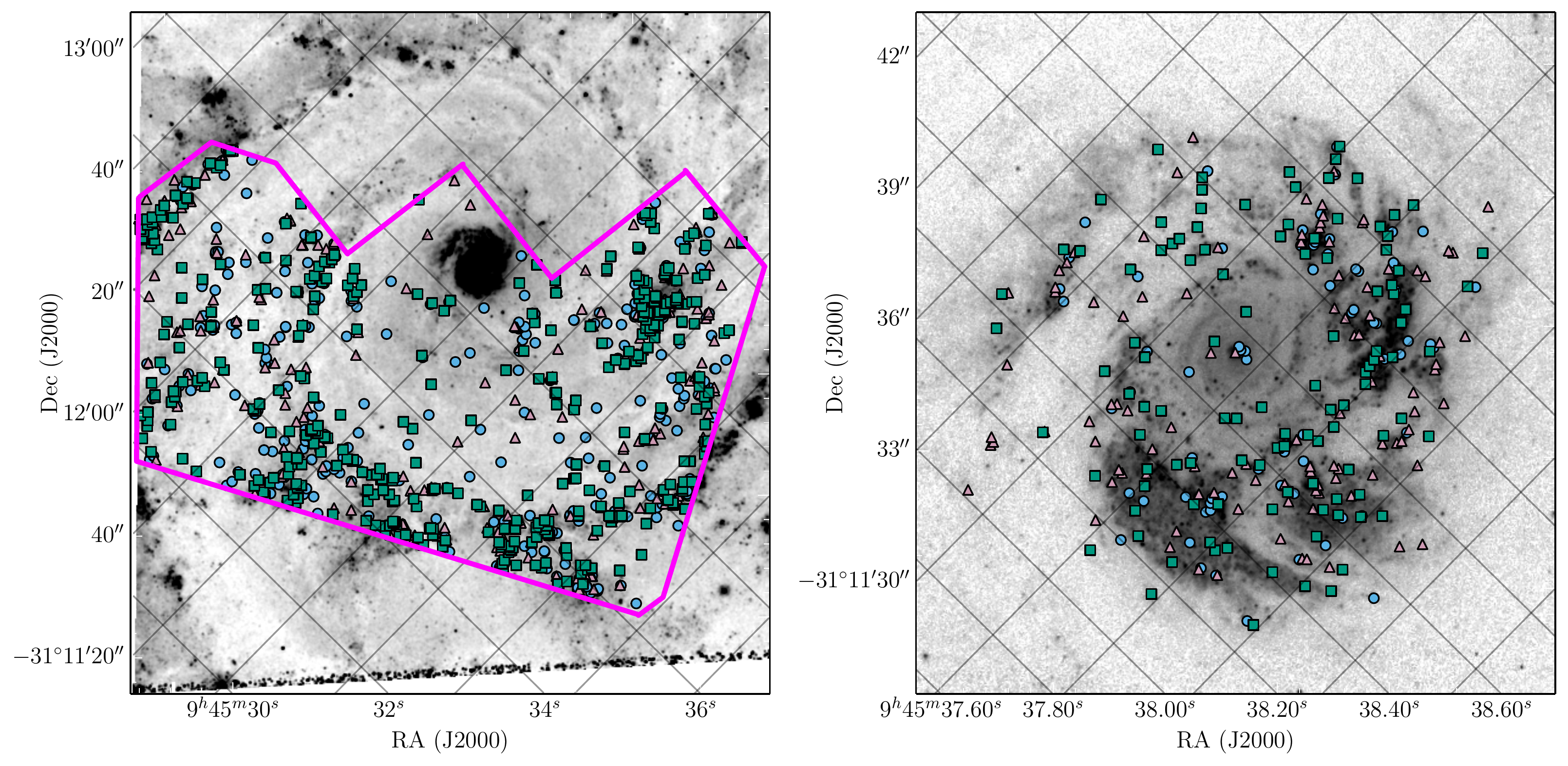}
\caption{Locations of (left) disk and (right) circumnuclear ring sources with $f_{UBVI}>1$ on the $U_{336}$ image.  The shape and color of the points depends on the $f_{UBVI}$ and $f_{\mathrm{vis}}$ flags assigned to the source. The blue circles are those with $f_{UBVI}=11$, the pink triangles have $f_{UBVI}=111$, and the green squares have $f_{UBVI}=111$ and $f_{\mathrm{vis}} = 1$ or $2$. North is oriented towards the bottom right and east is oriented towards the upper right. Grid lines of constant RA and Dec are overplotted on the image. \label{radecmap}}
\end{figure}

In Figure~\ref{radecmap}, we show the RA-Dec locations of sources with $f_{UBVI}>1$, separated by disk and ring sample, overplotted on the $U_{336}$ image. We perform our analysis only on those sources with $f_{UBVI} = 111$ and $f_{\mathrm{vis}} = 1$ or $2$. These sources constitute our high-confidence sample. There are 384 clusters in the disk sample and 110 in the circumnuclear ring sample.

\subsection{Extinction, H$\alpha$ Flagging, \& SED-fitting}
\label{halpha}

We took advantage of narrowband H$\alpha$ imaging of NGC~2997 from \cite{larsen1999a} to yield broad age and extinction constraints on the model grid used by our SED-fitting technique. We expect much of our sample to be subject to low amounts of extinction. This is because we detect and visually inspect our sources using the $U_{336}$ image. Reddening by dust increases dramatically at bluer wavelengths, and so any objects that are detectable across the $UBVI$ baseline should have relatively low extinctions. See Section~\ref{clfs} for further discussion of the effects of extinction on our CLFs.

Our continuum-subtracted H$\alpha$ image isolates nebular emission from HII regions from any stellar contribution to the H$\alpha$ flux. For every source in the disk sample with $f_{UBVI}=111$ and $f_{\mathrm{vis}}=1$ or $2$, we assigned a flag designating the absence or presence of H$\alpha$ at the cluster's location. Source confusion and crowding made it impossible to associate H$\alpha$ emission in the circumnuclear region with individual circumnuclear sources. The flags were assigned based on the following criteria:

\begin{itemize}
\item[0:] H$\alpha$ emission completely absent.
\item[1:] H$\alpha$ emission present and centered on the source.
\item[2:] H$\alpha$ emission present, but faint or not clearly associated with the source.
\item[99:] Circumnuclear ring source, not checked for H$\alpha$ emission.
\end{itemize}

According to self-consistent models which include both stellar and nebular continuum and emission (\textit{Yggdrasil}, \citealt{zackrisson2011}), we expect star clusters younger than 7~Myr to harbor enough young massive stars to ionize HII regions. Our detection limits, as described in Section~\ref{detlims}, prevent us from detecting very low mass clusters. It is therefore very unlikely that clusters without massive stars exist in our sample. Young clusters may also be partially embedded in their natal molecular cloud, so they may suffer from relatively large internal extinctions. At ages older than $\sim$7~Myr, we expect the majority of the natal, dusty material to have been expelled from the cluster environment, resulting in $A_{V}<1$~mag \citep[e.g.,][]{grosbol2009,grosbol2012}. However, we note that optical \textit{HST} surveys as well as resolved studies of young massive clusters in the Galaxy have suggested that clusters remain embedded for as little as 3~Myr or less \citep[e.g.,][]{longmore2014}. In any case, we expect older clusters to be associated with little to no H$\alpha$ emission. Finally, we note that the circumnuclear ring suffers from significant patchy extinction and non-local nebular emission not necessarily related to the cluster (see Figure~\ref{uvi-image}). We therefore cannot constrain the ages and extinctions of sources in the ring. Following these assumptions, the grid of models used in the SED fit is constrained as follows for each H$\alpha$ flag:

\begin{itemize}
\item[0:] Age constrained to 7 Myr to 14 Gyr. Internal extinction constrained to $E(B-V)\leq0.3$~mag.
\item[1:] Age constrained to 1 to 7 Myr. Internal extinction unconstrained, $0.0\leq E(B-V)\leq3.0$~mag.
\item[2:] Age unconstrained, 1~Myr to 14~Gyr. Internal extinction constrained to $E(B-V)\leq0.6$~mag\footnote{We choose this upper limit because the best-fit cluster masses and extinctions for these sources were found to be correlated above $E(B-V)\sim0.6$.}.
\item[99:] Age unconstrained, 1~Myr to 14~Gyr. Internal extinction unconstrained, $0.0\leq E(B-V)\leq3.0$~mag.
\end{itemize}

The spectral energy distributions (SEDs) of all sources with detections in at least three of the $UBVI$ filters were fit with \textit{Yggdrasil} models of three different metallicities, $Z=0.008,0.02,$ and $0.05$ \citep{zackrisson2011}. For each source, we select the model with the smallest reduced $\chi^2$ to assign the source's age, mass, and extinction. We refer the reader to \cite{konstantopoulos2013} and \cite{adamo2010a} for more detailed descriptions of the models and fitting procedure (see also Figure~7 in \cite{konstantopoulos2013} and Figures~A1~and~A2 in \citealt{adamo2012} for example SED fits). The $A_{V}$ distribution of our highest confidence sample ($f_{UBVI}=111$ and $f_{\mathrm{vis}}=1$ or 2) is similar to those found in other spiral galaxies and even starbursts (e.g., M51, \citealt{bastian2005}; the Antennae, \citealt{mengel2005}; Haro~11, \citealt{adamo2010a}), where at young ages ($\lesssim$10~Myr), $A_V$ ranges between 0 and $\sim$$2-3$~mag, and at older ages, $A_V$ drops to $\sim$0.2~mag. The age and mass distributions will be discussed in Sections~\ref{agemass} through \ref{ad}.

\subsection{Detection Limits}
\label{detlims}

We determine our detection limits by inspecting the CLF in each band (see Section~\ref{clfs} for a full description of the CLFs). We find that the cumulative distributions flatten out fainter than a certain brightness, deviating from a pure power-law shape. This flattening is likely due to a lack of clusters at the faint end of the CLF. In other words, we simply cannot detect a complete sample of clusters below a certain brightness in each filter with our data. The magnitude location of the turnover is defined to be the detection limit for that band. We chose this conservative definition of the detection limit because a typical completeness assessment is difficult for this dataset due to our combination of automatic and manual source selection. We find different detection limits for each band and sample (disk or circumnuclear ring) under question. The detection limits are listed in Table~\ref{detections}. 

\begin{table}
\begin{center}
\caption{Detection Limits \label{detections}}
\begin{tabular}{cc}
\tableline
Filter & Detection Limit \\
 & (mag) \\
\tableline
\multicolumn{2}{c}{disk} \\
\tableline
$U_{336}$ & 23.0 \\
$B_{450}$ & 23.9 \\
$V_{606}$ & 23.6 \\
$I_{\mathrm{WF}/814}$ & 23.3 \\
\tableline
\multicolumn{2}{c}{circumnuclear ring} \\
\tableline
$U_{336}$ & 22.1 \\
$B_{450}$ & 22.7 \\
$V_{555}$ & 22.5 \\
$I_{\mathrm{HRC}/814}$ & 22.0 \\
\tableline
\end{tabular}
\end{center}
\end{table}

\section{Properties of the Star Cluster Population of NGC~2997}
\label{clusterprop}

\subsection{Color-Color Diagrams}
\label{ccds}

\begin{figure}
\centering
\includegraphics[width=\textwidth]{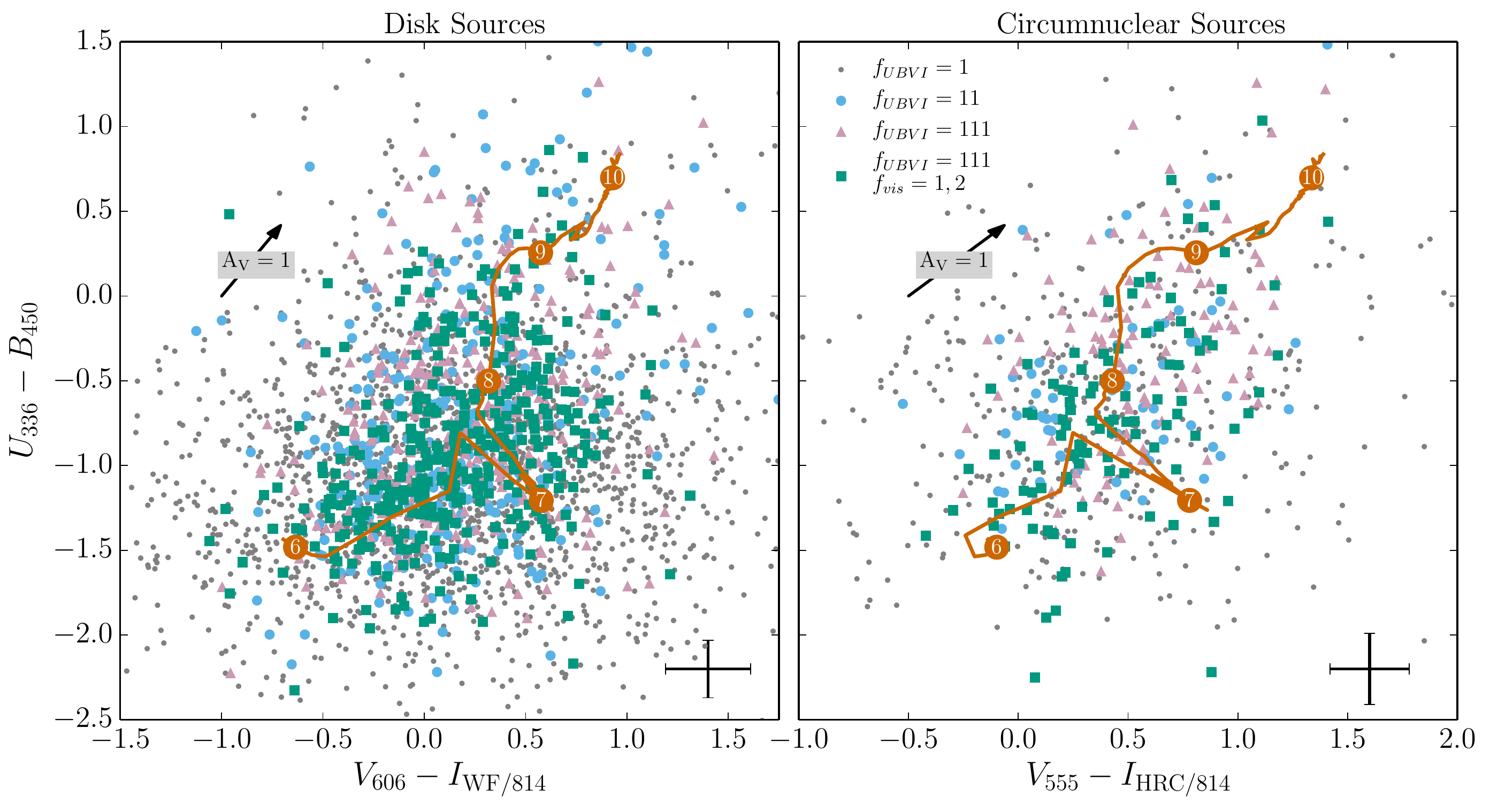}
\caption{Color-color diagram of disk (left) and circumnuclear ring (right) sources. The shape and color of the points depends on the $f_{UBVI}$ and $f_{\mathrm{vis}}$ flags assigned to the source. The gray dots are those with $f_{UBVI}=1$, the blue circles have $f_{UBVI}=11$, the pink triangles have $f_{UBVI}=111$, and the green squares have $f_{UBVI}=111$ and $f_{\mathrm{vis}} = 1$ or $2$. The orange line is the evolutionary track of a modeled SSP. The numbers on the track correspond to $\log(t_{\mathrm{age}})$. The black arrow corresponds to an $A_V = 1$.  \label{color-color}}
\end{figure}

In Figure~\ref{color-color}, we show the color-color diagrams of the sources in NGC~2997. On the left we plot $U_{336} - B_{450}$ vs. $V_{606} - I_{\mathrm{WF}/814}$ for the disk sources. On the right we plot $U_{336} - B_{450}$ vs. $V_{555} - I_{\mathrm{HRC}/814}$ for the circumnuclear sources. The color and shape of the data points corresponds to the flags assigned during the cluster selection process, and typical photometric errors are shown in the lower right corners. Our highest confidence sample, sources with $f_{UBVI}=111$ and $f_{\mathrm{vis}}=1$ or 2, are plotted as green squares. The solid orange lines are \textit{Yggdrasil} model tracks of solar metallicity \citep{zackrisson2011} with points denoting $\log(t_{\mathrm{age}})$ overplotted. 

As the selection criteria become more stringent ($f_{UBVI}$ increases), the scatter in color-color space clearly decreases. The points follow the evolutionary track well, suggesting the sample suffers from low amounts of extinction. In addition, a dearth of sources is seen at ages greater than $\sim$10$^8$~yr as compared to younger ages. Again, this likely due to our choice to use the $U_{336}$ image as the reference frame for cluster detection and analysis, i.e., the sample is luminosity-limited. Finally, we do not see significant differences in the color distribution between the disk and circumnuclear ring sources; both span similar ranges in color and age.

In NGC~4041, on the other hand, the cluster population of the inner field is bluer on average than that of the outer field, and younger clusters are more readily found in the outer field than the inner \citep{konstantopoulos2013}. These trends might be expected if a merger occurred recently, but could also be influenced by differing detection limits and crowding effects in the two fields. Still, it is noteworthy that the disk and circumnuclear ring in NGC~2997 lack a marked difference in cluster colors.

\subsection{Cluster Luminosity Functions}
\label{clfs}

Figure~\ref{CLF} shows the CLFs of our cluster samples. In this and all following figures, we plot only our highest confidence sample, those sources with $f_{UBVI}=111$ and $f_{\mathrm{vis}}=1$ or 2. The colors of the points correspond to the filter from which the brightness of each cluster is measured (\textit{U}: pink, \textit{B}: blue, \textit{V}: yellow, \textit{I}: orange), and the shapes correspond to the circumnuclear ring (triangles) and disk (circles) samples. We plot the cumulative distribution of the luminosity, so each point represents a single cluster. The vertical location of each point indicates the relative frequency with which a cluster of that magnitude appears in the sample. 

At the faint end of the distribution, the CLF flattens due to the instrumental detection limits, as discussed in Section~\ref{detlims}.  To the right of this turnover, we see a linear trend consistent with a power-law shape in log-log space. We used the \texttt{statpl} power-law fitting package to calculate the power-law index over the linear region of each CLF, only fitting those sources brighter than the detection limit \citep{maschberger2009, maschberger2012}. Using a modified maximum-likelihood fitting method, we find power-law indices of $-2.0$ to $-2.1$ for the disk in each filter. For the circumnuclear ring, we find power-law indices of $-2.1$ to  $-2.3$ in the optical $BVI$ filters. These values are consistent with CLF power-law indices reported in the literature for other nearby galaxies \citep[e.g.,][]{whitmore1999,larsen2002,bastian2012a,konstantopoulos2013}. The circumnuclear ring $U_{336}$ CLF is best fit with a rather shallow power-law index of $-1.7$, which is likely due to the effects of crowding and source confusion removing faint clusters from the sample.

In their near-infrared study of cluster complexes in NGC~2997, \cite{grosbol2012} find power-law indices of $-2.29$ for sources younger than $\sim$7~Myr and $-2.49$ for older sources after correcting for extinction. These values are steeper than those we find in the optical, which is expected for luminosity functions derived from redder bands \citep{gieles2010}. As discussed in Section~\ref{halpha}, our cluster sample suffers from rather small amounts of extinction. However, to ensure that extinction is not affecting our CLFs, we correct each cluster for its best-fit extinction in each band and refit the power-laws. We find slightly shallower best-fit indices, but they remain consistent with the indices of the uncorrected CLFs within the errors, which agrees with the findings of \cite{larsen2002}. We also note that the objects \cite{grosbol2012} detects in NGC~2997 are likely to be complexes of clusters, the luminosity functions of which may behave differently than those for individual clusters.

All of the disk CLFs and the $U_{336}$ circumnuclear ring CLF are steeper at the bright end of the distribution than at the faint end (to the right of the detection limit turnover). Bends such as these are expected if there is an underlying truncation in the cluster mass function \citep{gieles2006b}. The \texttt{statpl} software also performs goodness-of-fit and truncation tests, which we use to test the significance of the bends we observe. For the disk sample, the goodness-of-fit tests suggest that the CLFs are not entirely consistent with a power-law, but also are not truncated. Only the $I_{\mathrm{WF}/814}$ CLF is somewhat consistent with a truncation at a cluster magnitude of 19.0. The circumnuclear ring sample CLFs are consistent with power-laws, except for the $U_{336}$ CLF, which has a possible truncation at a cluster magnitude of 16.8. The $U_{336}$ CLF is also significantly shallower than the others, which could be due our detection limits preferentially removing faint clusters from the sample. We expect to find bends in the CLF, not truncations, so we take these possible truncations as evidence for deviations from pure power-law behavior that hint at a truncation in the mass function.

We also observe a slight steepening of the CLFs in the redder bands, especially in the circumnuclear ring. \cite{gieles2010} showed that if cluster disruption is negligible, then the CLF will be steeper in redder bands simply due to evolutionary fading of the cluster population. A low rate of cluster disruption, at least in the circumnuclear ring, could cause the CLFs to steepen with increasing wavelength. We address cluster disruption again in Section~\ref{ad}.

\begin{figure}
\centering
\includegraphics[scale=0.6]{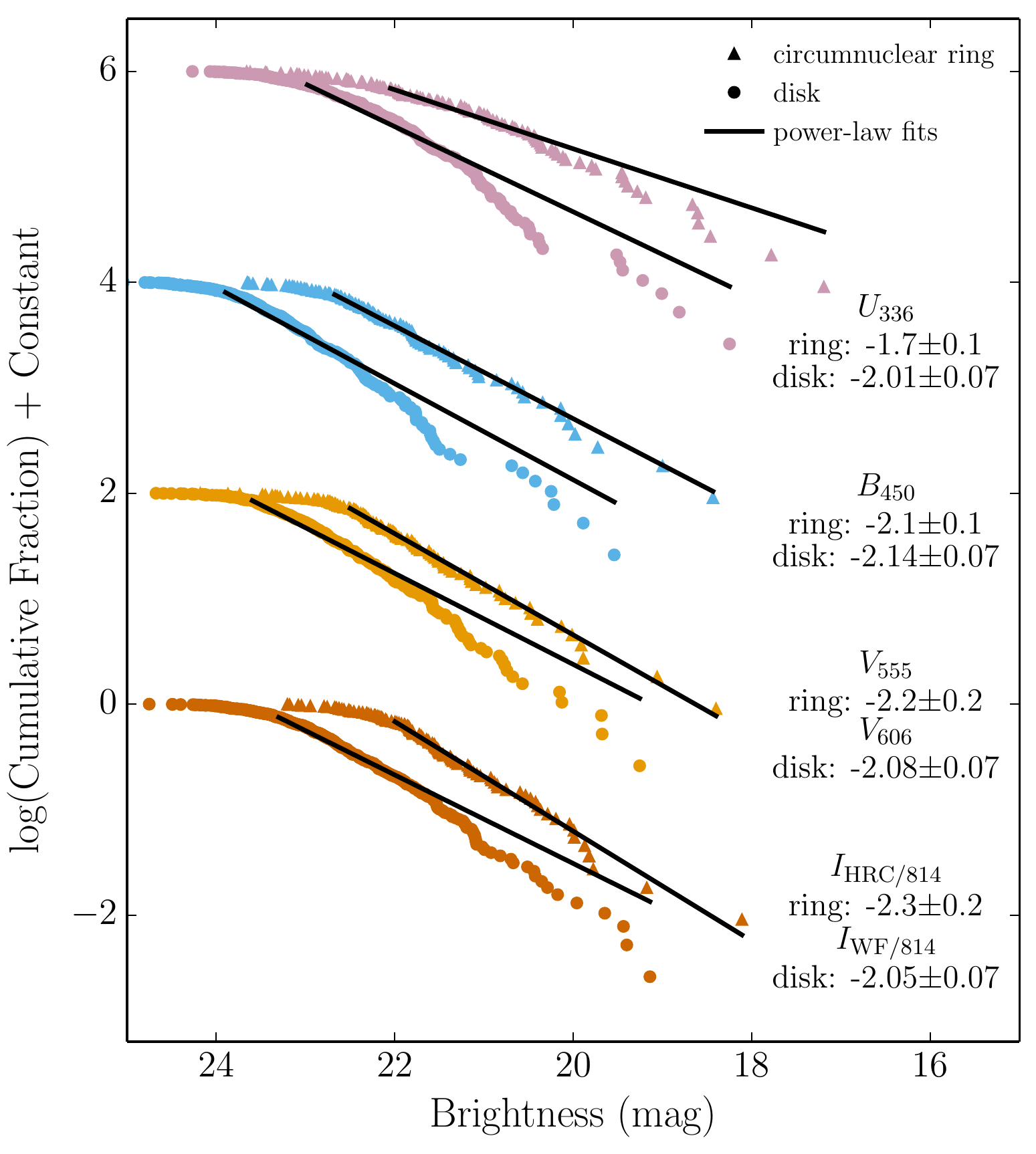}
\caption{Cumulative luminosity functions for clusters with $f_{UBVI} = 111$ and $f_{\mathrm{vis}} =$ 1 or 2. The disk and circumnuclear sources are plotted separately for each detector-filter combination. The black lines represent maximum-likelihood fits to the CLFs, and the power-law indices recovered from the fits are labeled to the right of the respective LFs. \label{CLF}}
\end{figure}

\subsection{Best-fit Ages and Masses}
\label{agemass}

\begin{figure}
\centering
\includegraphics[scale=0.6]{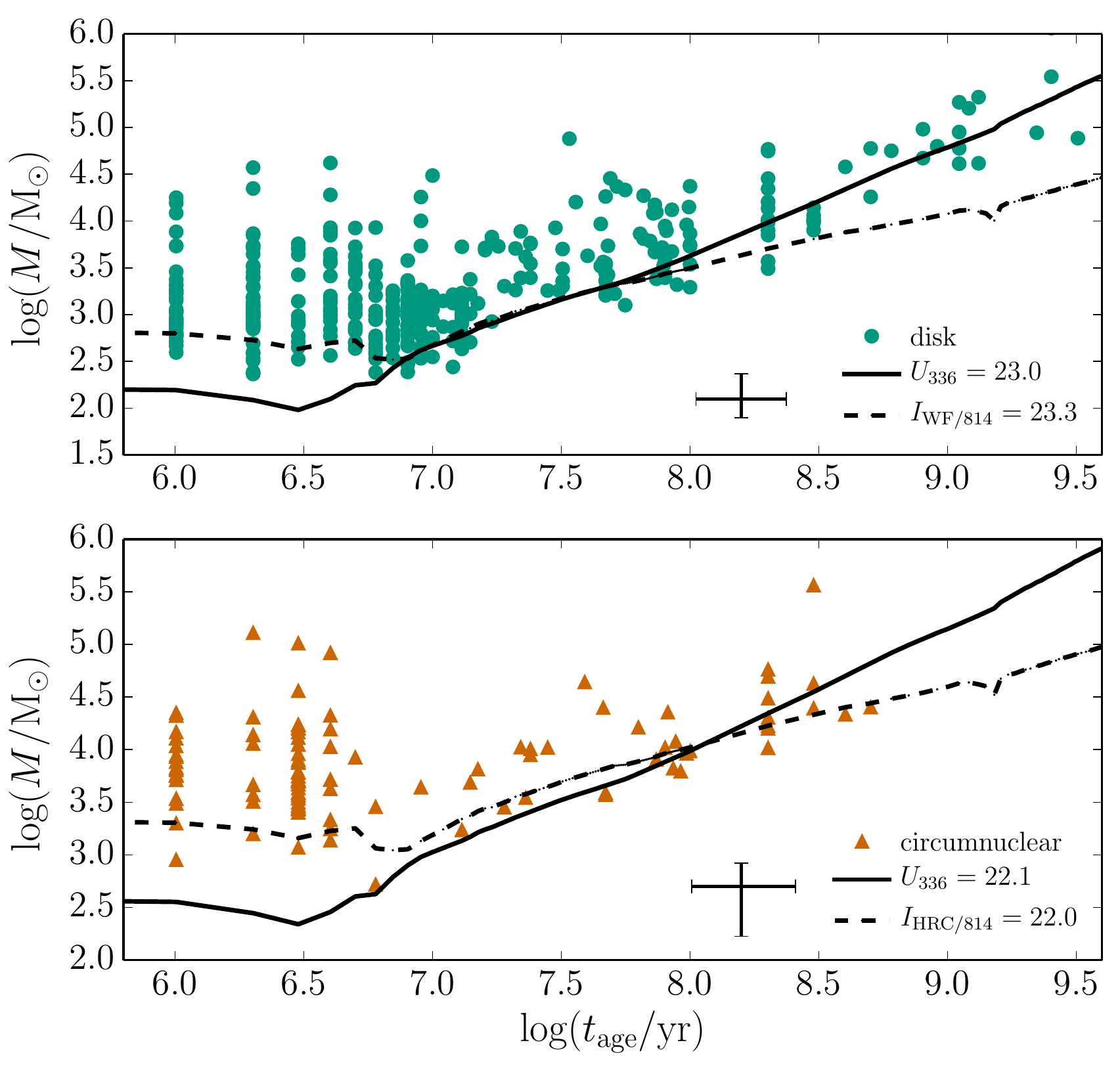}
\caption{Age-mass plot of disk (top) and circumnuclear (bottom) clusters with $f_{UBVI} = 111$ and $f_{\mathrm{vis}} =$ 1 or 2. The black lines are evolutionary tracks of modeled SSPs corresponding to the $U_{336}$ (solid), $I_{\mathrm{WF}/814}$ (top frame, dashed), $I_{\mathrm{HRC}/814}$ (bottom frame, dashed) detection limits. Representative error bars are plotted at the bottom right of each frame. \label{age-mass}}
\end{figure}

Figure~\ref{age-mass} shows the ages and masses of the clusters in the disk (top, green circles) circumnuclear ring (bottom, orange triangles). Typical errors are plotted near the bottom of each frame and the solid and dashed lines represent the estimated detection limits in $U_{336}$ and $I_{\mathrm{WF}/814}$ or $I_{\mathrm{HRC}/814}$, respectively. These limits show how massive a cluster of a given age must be in order to be reliably detected. Linear features in the data are due to the timestep resolution of our SSP models. 

Small differences are seen between the two samples. Several disk clusters have ages $>$400~Myr, whereas in the circumnuclear ring, there are few to none. Source confusion and crowding in the circumnuclear ring may be preventing detection of older, faint clusters in the ring. In addition, young clusters in the ring have higher masses than those of the same age in the disk. Our limited $UBVI$ footprint of the disk likely explains the lack of high mass clusters in the disk sample. Ground-based studies with wider spatial coverage have found many objects of higher mass than we find here \citep{larsen1999a, grosbol2006}, though these studies may include cluster complexes in their samples due to resolution effects. A larger sample of clusters, which would be detected with more extensive $UBVI$ coverage of the galaxy's disk, would allow us to better populate the age and mass distributions. 

\subsection{Cluster Mass Function}
\label{cmf}

\begin{figure}
\centering
\includegraphics[scale=0.6]{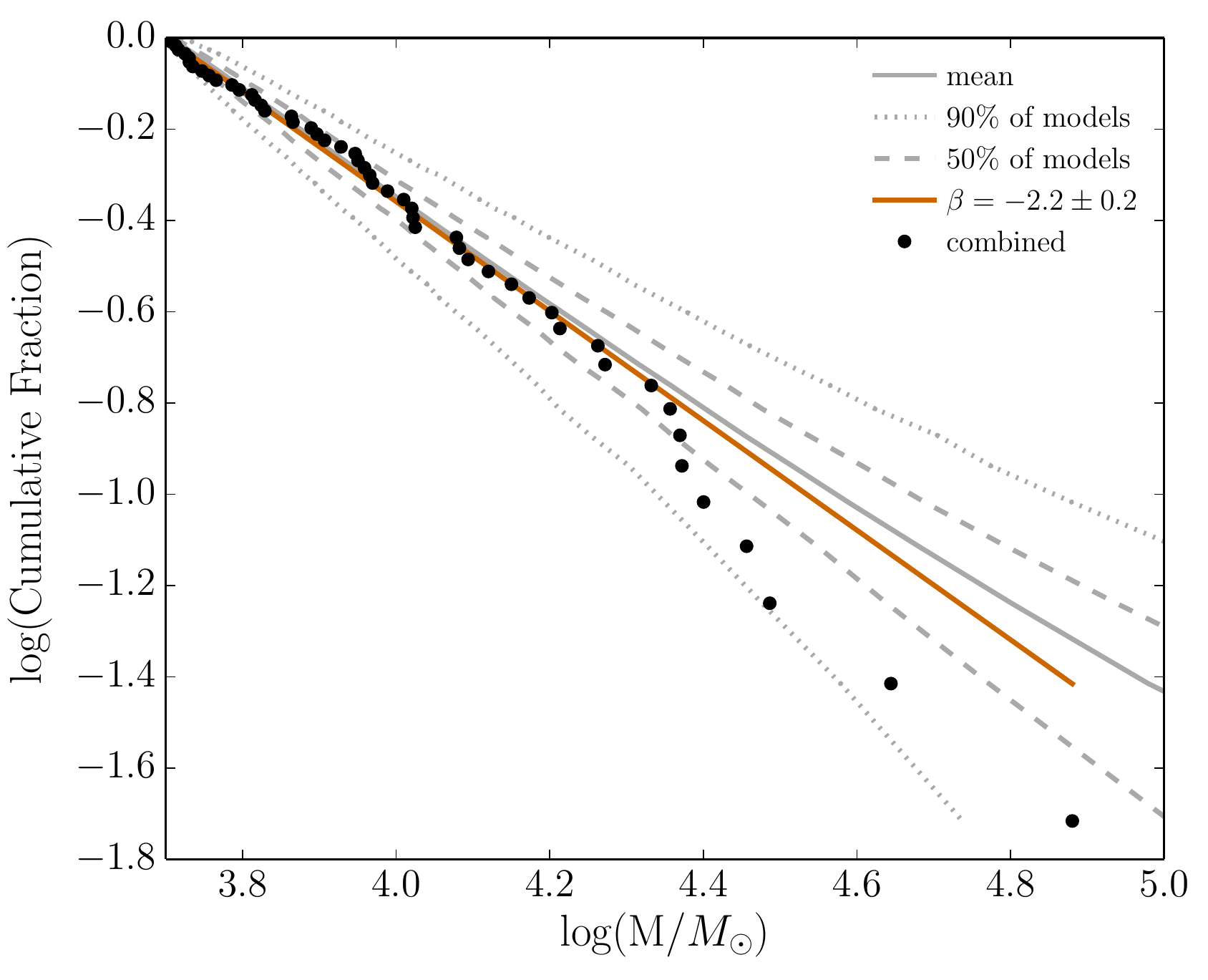}
\caption{Cumulative mass function (MF) for clusters with $f_{UBVI} = 111$, $f_{\mathrm{vis}} =$ 1 or 2, $M > 5\times10^3$~M$_{\odot}$, and $10 < t_{\mathrm{age}} < 100$~Myr. The circumnuclear and disk cluster samples are combined in this plot. The orange line represents a modified maximum-likelihood power-law fit to the data, resulting in an index of $-2.2\pm0.2$. The gray lines represent the mean (solid), 50th percentile range (dashed), and 90th percentile range (dotted) of 2000 Monte Carlo simulations of cluster populations generated by sampling a power-law MF with index $-2.2$. \label{CMF}}
\end{figure}

From Figures~\ref{color-color}, \ref{CLF}, and \ref{age-mass}, we find that the disk and circumnuclear ring clusters do not appear to have significantly different properties. We therefore assume they are not separate populations. In Figure~\ref{CMF}, we combine the two samples and plot the cumulative distribution of the resulting mass function. We remove sources younger than 10~Myr to avoid contamination by possible unbound stellar associations masquerading as clusters at the distance of NGC~2997. We also remove clusters older than 100~Myr to avoid the effects of evolutionary fading, which causes the majority of clusters to drop below our detection limits.

We include only sources more massive than $5\times10^3$~M$_{\odot}$ to ensure our sample is complete to ages of 100~Myr (see the intersection of this mass cut and the detection limits in Figure~\ref{age-mass}). This mass cut also minimizes the effects of assuming the stellar initial mass function (IMF) is continuously sampled in star clusters of all masses, while in reality the finite number of stars in each cluster means the IMF is stochastically sampled. Such an assumption can strongly affect the ages and masses derived from broad-band magnitudes of clusters, but this effect diminishes at high masses \citep[e.g.,][]{popescu2010,silvavilla2011}. In addition, the overall age and mass distributions of cluster populations are not strongly affected by the sampling method assumed, unlike the properties of individual clusters \citep{fouesneau2010}. 

After the age and mass cuts, we are left with a total of 52 disk and circumnuclear clusters. Again we use the \texttt{statpl} power-law fitting package to calculate the power-law index of the CMF \citep{maschberger2012}. Our data are best fit with a power-law of index $-2.2\pm0.2$, represented by the orange line in Figure~\ref{CMF}. This power-law index is consistent with others from the literature for nearby galaxies \citep[e.g.,][]{zhang1999,bik2003,larsen2009,bastian2012a,konstantopoulos2013}. If we restrict our sample to $f_{\mathrm{vis}}=1$ sources only (the `good' sources from the visual inspection), the best-fit power-law index does not change, and the overall shape of the CMF is retained.

The data appear to deviate from a pure power-law above $\sim$$2.5\times10^{4}$~M$_{\odot}$. To determine whether this is evidence of a truncation in the CMF, we generated 2000 cluster populations using a Monte Carlo method. Each population contains the same number of clusters as our data. We randomly sampled a power-law mass function with index $-2.2$ to generate each model population's mass function. The mean of the models is shown as a solid line, while 50\% and 90\% of the models fall within the dashed and dotted lines, respectively. At no point do the data fall outside the 90\% range of the models, suggesting that the deviation is not a statistically significant truncation. To further test this result, we run truncation tests provided by the \texttt{statpl} package on our data. The exceedance test, which is the strongest test for truncation, finds that the CMF is not truncated to a significance level of 5\%. 

Although studies of other nearby galaxies have suggested the existence of a truncation in the CMF, it is unsurprising that we do not observe such a truncation in NGC~2997. In spiral galaxies, the average truncation mass is $\sim$$2\times10^5$~M$_{\odot}$ \citep{larsen2009}, which is larger than the most massive cluster included in our CMF. Indeed, the spectroscopic study of NGC~2997's CMF by \cite{bastian2012b}, using the method of \cite{larsen2009}, found that a Schecter function best described their data, with a truncation mass of $5-6\times10^{5}$~M$_{\odot}$. Few of the clusters included in \cite{bastian2012b} are present in our $UBVI$ footprint, though their ground-based data may lead to an overestimate of the actual luminosities (and therefore masses). It is also possible our limited $UBVI$ footprint is preventing us from fully probing the CMF of this galaxy, i.e., we are missing star clusters in the high mass regime.

\subsection{Cluster Age Distribution and Disruption}
\label{ad}

\begin{figure}
\centering
\includegraphics[scale=0.6]{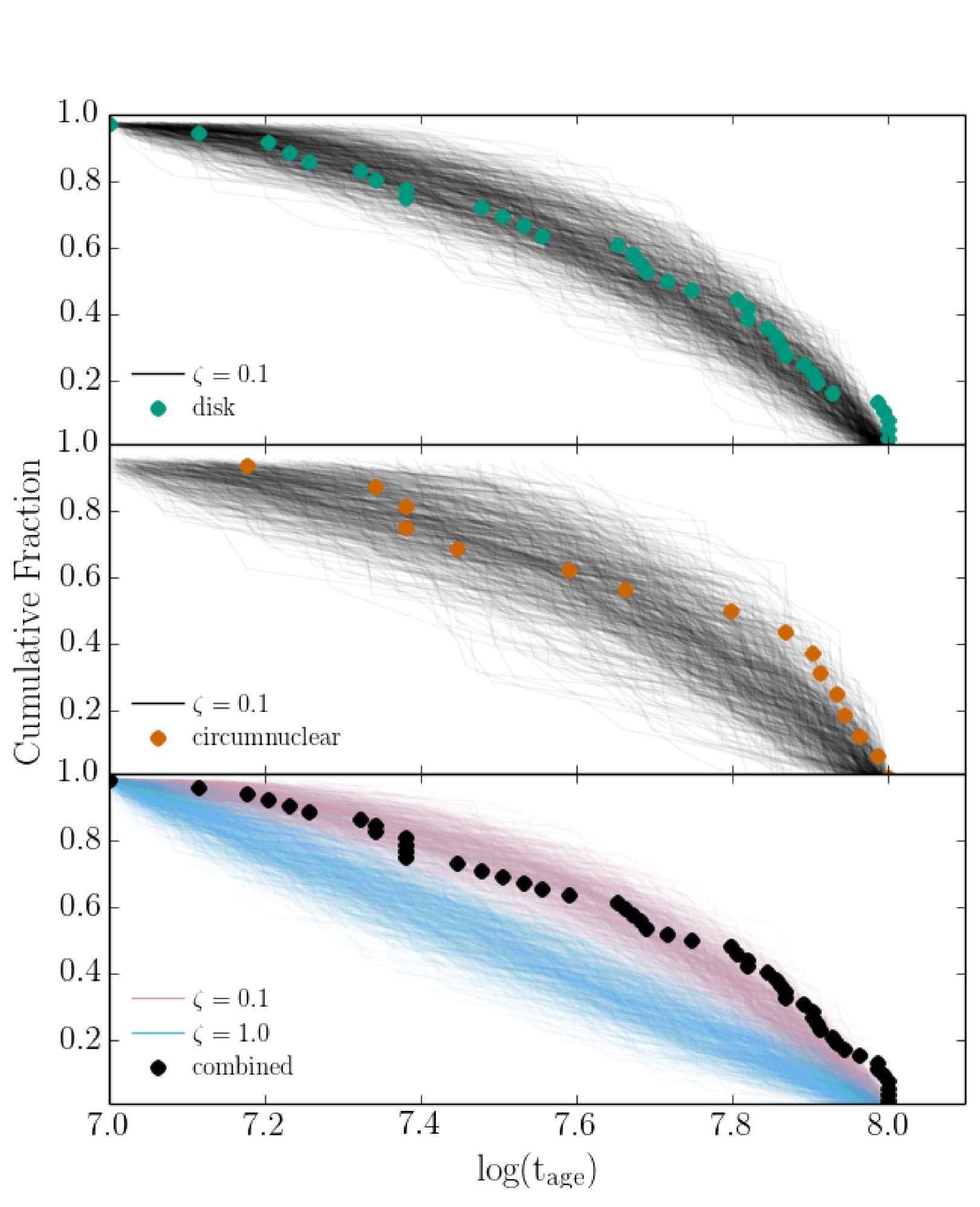}
\caption{Cluster age distribution plotted as a cumulative fraction for the disk clusters (top), circumnuclear ring clusters (middle), and the combined sample (bottom). Only clusters with  $f_{UBVI} = 111$, $f_{\mathrm{vis}} =$ 1 or 2, $M > 5\times10^3$~M$_{\odot}$, and $10 < t_{\mathrm{age}} < 100$~Myr are plotted here. Each thin line is a Monte Carlo simulation of a cluster population subject to cluster disruption that takes the form of a power-law, $t_{\mathrm{age}}^{\zeta}$. In the top two plots, $\zeta = -0.1$. In the bottom plot, the pink lines are for $\zeta = -0.1$ and the blue lines are for $\zeta=-1.0$.  \label{dNdt}}
\end{figure}

In Figure~\ref{dNdt}, we plot the age distributions of the disk (top), circumnuclear ring (middle), and combined (bottom) samples using cumulative fractions. Only those clusters with ages between 10 and 100~Myr and masses higher than $5\times10^3$ M$_{\odot}$ are plotted here, as in Figure~\ref{CMF}. These age and mass cuts ensure our sample is mass-limited. If our sample were luminosity-limited, it would be biased towards younger clusters, artificially steepening the age distribution. With a mass-limited sample, we can assume that only cluster formation and disruption are setting the shape of the age distributions \citep{gieles2007}. Because NGC~2997 appears rather undisturbed, we also assume that the cluster formation rate has been roughly constant over the last 100~Myr. Therefore, it is likely that cluster disruption is solely responsible for determining the shape of the cluster age distributions.

We choose to compare the age distribution to a simple power-law of the form $dN/dt \propto t_{\mathrm{age}}^{\zeta}$, as has been done in several recent studies \citep[e.g.,][]{bastian2005, fall2005, gieles2007, chandar2010, portegieszwart2010, silvavilla2011, silvavilla2014}. The value of the power-law index $\zeta$ has been observed to range from $\sim$$-0.1$ to $\sim$$ -1.0$ in nearby galaxies. \cite{grosbol2013} find a $\zeta=-1.4$ for cluster complexes in NGC~2997, and suggest this is because complexes disrupt into individual clusters at a high rate. We note that no age or mass cuts have been applied to their sample, possibly biasing the age distribution towards younger objects and steepening the power-law trend.

We generate cluster populations using a Monte Carlo method to test the shape of the age distribution. For the top and middle panels of Figure~\ref{dNdt}, we create 500 cluster populations by randomly sampling a power-law mass function of index $-2.2$. We randomly assign an age to each cluster along with a probability of survival determined by the power-law distribution $dN/dt \propto t_{\mathrm{age}}^{-0.1}$. If the probability of survival is greater than a randomly generated number between 0 and 1, the cluster survives. This method treats clusters of all masses equally, i.e., cluster disruption is implemented in a mass-independent manner. We apply the same age and mass cuts to each cluster population as were applied to the data. Each cluster population generated in this way contains a different number of surviving clusters. By adjusting the initial number of clusters generated, we ensure that the average number of surviving clusters over all 500 cluster populations matches the number of clusters in our data. 

Each cluster population is plotted as a thin black line with 95\% transparency. The data appear to follow the model cluster populations well, although the scatter in the models is large in both cases due to small sample sizes. A larger scatter in the model age distributions is seen in the middle panel because the circumnuclear ring sample contains 16 clusters, whereas the disk sample contains 36 clusters.

For the bottom panel, we generate two sets of cluster populations using the same method as above. The first set is subject to mild disruption, with a power-law index of $\zeta = -0.1$, and is plotted as pink lines. The second set is subject to strong disruption, with $\zeta=-1.0$, and is plotted as blue lines. We find that the data agree more clearly with the $\zeta=-0.1$ set of models. If we restrict all three age distributions to $f_{\mathrm{vis}}=1$ sources only, we find that the overall shape does not change significantly, so the data still agree with the $\zeta=-0.1$ set of models.

Given the scatter in the models, it is possible that the age distributions are also consistent with steeper (more negative) values of $\zeta$, but it is clear from the bottom panel of Figure~\ref{dNdt} that our data are not consistent with values as steep as $\zeta=-1.0$. We therefore conclude that the clusters in NGC~2997 have not been subject to a significant amount of disruption in the age range considered, 10 to 100~Myr. 

We find that the circumnuclear ring and disk samples appear to have similar, shallow values of $\zeta$, suggesting that the rate of cluster disruption does not vary as a function of environment. However, we cannot distinguish small differences in $\zeta$ between the two samples because of the large scatter caused by small number statistics. Therefore, we can neither confirm nor contradict recent studies that find the power-law index of the age distribution becomes steeper in regions of higher density \citep{bastian2012a,silvavilla2014}. Again, better $UBVI$ coverage of NGC~2997 would provide a larger cluster sample with which to probe this issue.

\subsection{Efficiency of Cluster Formation as a Function of the Environment}
\label{cfe}

\begin{figure}
\centering
\includegraphics[scale=0.6]{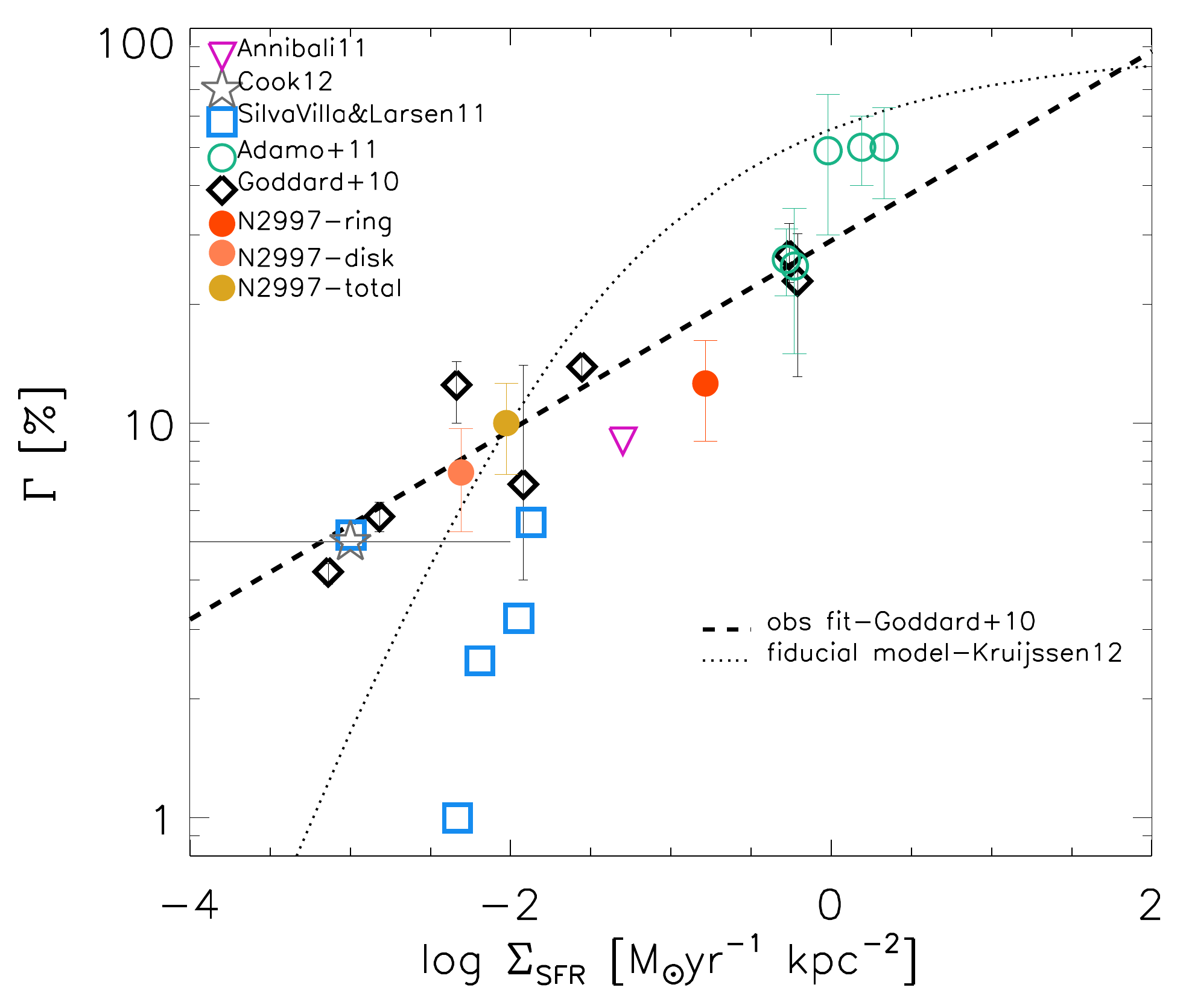}
\caption{Cluster formation efficiency ($\Gamma$) vs SFR surface density ($\Sigma_{\mathrm{SFR}}$) for several different galaxies and regions within galaxies. Our values for NGC~2997 are plotted as yellow, orange, and red points for the total $UBVI$ footprint, disk region, and circumnuclear ring, respectively. The other values are taken from the literature. The dashed line is a power-law fit to the black points from \cite{goddard2010}. The dotted line is a model trend from \cite{kruijssen2012a}.  \label{gamma}}
\end{figure}

In Figure~\ref{gamma}, we plot our values of the cluster formation efficiency, $\Gamma$, as a function of star formation rate density, $\Sigma_{\mathrm{SFR}}$, along with values from the literature. \cite{bastian2008b} defined $\Gamma$ to be the cluster formation rate (CFR) divided by the SFR in a given region, with both the CFR and SFR calculated for the same period of time. The CFR is defined as the total mass formed in clusters over a given time period, or $M_{\mathrm{tot,cl}}/\Delta t$. To calculate the CFR, we first sum the masses of clusters between 1 and 100~Myr in age and above a mass of 10$^4$ ~M$_{\odot}$. In doing so, we assume that we have detected all the clusters above this mass limit and within this age range in our $UBVI$ footprint. We choose this age range because it matches the age range over which the SFR calculated in the next paragraph applies. To estimate the total mass in clusters, we find the missing fraction in low mass clusters ($10^2 \leq M_{\mathrm{cl}} < 10^4$~M$_{\odot}$) by integrating over a power-law mass function of index $-2$ and assuming that the total mass in high mass clusters ($10^4 \leq M_{\mathrm{cl}} < 10^7$~M$_{\odot}$) is known. Over the last 100~Myr, we find the CFR to be 0.008~M$_{\odot}$~yr$^{-1}$ in the disk and 0.014~M$_{\odot}$~yr$^{-1}$ in the circumnuclear ring. For the entire $UBVI$ footprint, we find a CFR of 0.022~M$_{\odot}$~yr$^{-1}$.

We calculate the total SFR in both the disk and circumnuclear region by summing FUV and 24~$\mu$m SFRs calculated from archival \textit{GALEX} and \textit{Spitzer} imaging. We use the FUV flux to estimate the SFR from massive stars unobscured by dust, and the 24~$\mu$m flux to estimate the SFR from dust heated by highly obscured massive stars. Both trace star formation with age $\leq$100 Myr. 

We use \texttt{polyphot} to measure the FUV and 24~$\mu$m in the $UBVI$ footprint and \texttt{apphot} to measure them in the circumnuclear ring using an aperture of radius 10'' (460~pc). We correct the FUV flux for foreground reddening and convert both fluxes to luminosities. Finally, we use the calibrations in Table~1 of \cite{kennicutt2012} to calculate the SFRs. We find that the disk has SFR$_{\mathrm{FUV}} = 0.04$~M$_{\odot}$~yr$^{-1}$ and SFR$_{24\mu\mathrm{m}} = 0.068$~M$_{\odot}$~yr$^{-1}$, for a total of 0.11~M$_{\odot}$~yr$^{-1}$. The circumnuclear ring has SFR$_{\mathrm{FUV}} = 0.01$~M$_{\odot}$~yr$^{-1}$ and SFR$_{24\mu\mathrm{m}} = 0.10$~M$_{\odot}$~yr$^{-1}$, for a total of 0.11~M$_{\odot}$~yr$^{-1}$. The disk and circumnuclear ring therefore have the same total SFR, but widely different $\Sigma_{\mathrm{SFR}}$, $0.0049$~M$_{\odot}$~yr$^{-1}$~kpc$^{-2}$ for the disk and $0.164$~M$_{\odot}$~yr$^{-1}$~kpc$^{-2}$ for the ring.

Our final values for $\Gamma$ are $7\pm2$\% for the disk, $12\pm4$\% for the circumnuclear ring, and $10\pm3$\% for the entire $UBVI$ footprint. To calculate the errors on our $\Gamma$ values, we assume a 20\% error on the SFRs. For the CFRs, we fold in the Poissonian error on the number of clusters and 0.2 dex uncertainties on cluster age and mass into 1000 Monte Carlo realizations of a cluster population. The final error on $\Gamma$ for each region is calculated by propagating the error on the CFR and SFR. If we use only $f_{\mathrm{vis}}=1$ sources to calculate the CFRs, our $\Gamma$ values decrease slightly, but remain consistent with those of the total sample within the errors.

Our values for $\Gamma$ are plotted in Figure~\ref{gamma} as yellow, orange, and red points for the total $UBVI$ footprint, disk region, and circumnuclear ring, respectively. The other points are values taken from the literature. The dashed line is a power-law fit to the black points from \cite{goddard2010}. The dotted line is a model trend of a `typical' disk galaxy from \citep{kruijssen2012a}. The steep increase of $\Gamma$ in this model is driven by increasing gas surface density, which allows gravitationally bound stellar systems to naturally form in the high-density, high SFE regions of the ISM. The turnover at $\Gamma\sim70\%$ is due to tidal disruption of young star-forming regions by neighboring gas clouds, which only has a strong effect at very high gas surface density. Our values are consistent with those from the literature, and follow the observed trend wherein environments of higher SFR surface density form a higher percentage of bound star clusters (as opposed to unbound associations or other forms of diffuse star formation). Our total footprint and disk region points agree very well with the Kruijssen model trend, but the circumnuclear ring is offset, suggesting its ISM properties may not match those of the rest of the galaxy.

For our two regions within NGC~2997, we see a possible environmental dependence in $\Gamma$; the circumnuclear ring has a slightly higher value than that of the rest of the disk. However, the $\Gamma$ values are consistent with each other within the errors, so again, a larger cluster sample would allow us to probe this result more fully. If true, this would agree with the result from \cite{silvavilla2013}, who found that $\Gamma$ decreased as a function of galactocentric radius for four radial bins in M83. This implies that the fraction of stars formed in bound clusters depends on local conditions within the host galaxy, and also correlates with galaxy-wide averages of those conditions.

\section{Conclusions}
\label{conclusions}

We have analyzed the star cluster population of NGC~2997 using $U$-band imaging obtained by SHUCS along with archival $BVI$ imaging from $HST$. Our conservative cluster selection criteria allowed us to select a high-confidence sample of star clusters in the circumnuclear ring and disk. We obtained ages and masses for each cluster by fitting their SEDs with SSP models. Finally, we have presented and interpreted the color, luminosity, mass, and age distributions of our cluster sample. The main results from this study are summarized as follows:

\begin{enumerate}
\item Ample spatial coverage across the $UBVI$ baseline is important for studies of star cluster populations in giant star-forming galaxies. Small $UBVI$ footprints can hamper attempts to fully probe the physical characteristics of cluster populations, although, as we show in this study, small sample sizes can nonetheless provide insight into these properties.
\item Conservative selection criteria are essential to building a bona fide cluster sample with which to probe the physical properties of the population. Selection critieria should include a combination of quantitative measurements of concentration and visual inspection for circular symmetry and extendedness. This minimizes contamination by individual stars and unbound stellar associations, which can bias interpretation of the data. Our selection criteria provide a high-confidence sample of 384 clusters in the disk and 110 in the circumnuclear ring. To study the physical properties of the cluster population, we impose further age and mass cuts on the high-confidence sample, leaving us with 36 clusters in the disk and 16 in the circumnuclear ring. These cuts ensure the sample is mass-limited to an age of 100~Myr, minimizing the effects of evolutionary fading at old ages, unbound associations at young ages, and stochastic sampling of the stellar IMF at low masses.
\item The cluster luminosity functions consist of a flat, incomplete part at the faint end and a linear (power-law) trend at the bright end. We fit the linear regimes and find power-law indices of $-1.7$ to $-2.3$, consistent with other indices in the literature. We find several of the CLFs deviate from pure power-law shape at the bright end, hinting at a high-mass truncation in the cluster mass function. We also note that the redder filters appear to have steeper CLFs, which is expected from evolutionary fading of star clusters.
\item The cluster mass function of the combined sample is consistent with a power-law of index $-2.2\pm0.2$. Using Monte Carlo simulations and goodness-of-fit tests, we find a deviation from a pure power-law shape at $\sim$2.5$\times$10$^{4}$~M$_{\odot}$, but no truncation. The lack of 10 to 100~Myr old clusters with masses above 5000~M$_{\odot}$ in our sample suggests our $UBVI$ footprint is simply too small to properly sample the highest cluster mass bins at young ages.
\item The cluster age distributions are consistent with a low rate of disruption, $\zeta\sim-0.1$, assuming the cluster formation rate has been roughly constant over the last 100~Myr. Small sample sizes prevent us from finding different values of $\zeta$ for the disk and circumnuclear ring. We therefore neither support nor refute recent results which show steeper age distributions for smaller galactocentric radii. However, our data do support the view that the age distribution of this mass-limited cluster sample is significantly shallower than $\zeta=-1.0$.
\item We calculate the cluster formation efficiency, $\Gamma$, over the last 100~Myr. We find $\Gamma$ is $7\pm2$\% for the disk, $12\pm4$\% for the circumnuclear ring, and $10\pm3$\% for the entire $UBVI$ footprint. These values are consistent with the literature in regions with similar values of $\Sigma_{\mathrm{SFR}}$. We see a possible environmental dependence in our $\Gamma$ values, which would support recent results that $\Gamma$ decreases with increasing galactocentric radius.
\end{enumerate}

\section{Acknowledgements}

Support for this work was provided by NASA through grant No. HST-GO-12229.01-A from the Space Telescope Science Institute, which is operated by AURA, Inc., under NASA contract NAS5-26555. This research has made use of the NASA/IPAC Extragalactic Database (NED) which is operated by the Jet Propulsion Laboratory, California Institute of Technology, under contract with the National Aeronautics and Space Administration. J.E. Ryon gratefully acknowledges the support of the National Space Grant College and Fellowship Program and the Wisconsin Space Grant Consortium. E. Silva-Villa is a postdoctoral fellow supported by the Centre de Recherche en Astrophysique du Qu\'ebec (CRAQ). I. S. Konstantopoulos is the recipient of a John Stocker Research Fellowship from the Science and Industry Endowment Fund (Australia). E. Zackrisson acknowledges funding from the Swedish National Space Board, the Swedish Research Council and the Wenner-Gren Foundations.

\bibliography{ngc2997}

\end{document}